\documentclass[aps,twocolumn,superscriptaddress,showpacs]{revtex4}

\usepackage[dvips]{graphicx}
\usepackage{amssymb}
\usepackage{color}
\newcommand{\nc}{\newcommand}		
\nc{\vc}[1]	{\mbox{\boldmath $#1$}}	
\nc{\del}       {\partial}              
\nc{\bra}       {\langle}               
\nc{\ket}       {\rangle}               
\nc{\bras}[1]   {\langle #1|}           
\nc{\kets}[1]   {|#1\rangle}            
\nc{\mapleft}[1]{			
 \smash{\mathop{\,			%
  \hbox to 1.5cm{\rightarrowfill}\, }\limits_{#1}}}
\nc{\beq}     {\begin{eqnarray}}
\nc{\eeq}    {\end{eqnarray}}
\nc{\nn}   {\\\nonumber}
\nc{\dl} {\delta}
\nc{\tht} {\theta}
\nc{\sig} {\sigma}
\nc{\Dl} {\Delta}
\nc{\Sig} {\Sigma}
\nc{\dg}  {\dagger}
\nc{\ti} {\tilde}
\nc{\lm} {\lambda}
\nc{\bx}     {\bold x}
\nc{\br} {\bold r}
\nc{\fra}     {\frac{1}{2}}
\nc{\AMD}{{\rm AMD}}
\nc{\red}[1]    {\textcolor{red}{#1}}  
\nc{\blue}[1]   {\textcolor{blue}{#1}}  
\nc{\green}[1]   {\textcolor{green}{#1}}  
\nc{\mydraft}	{\setlength{\topmargin}{-1.5cm}}
\mydraft

\begin{document}
\title{Tensor-optimized antisymmetrized molecular dynamics\\ as a successive variational method in nuclear many-body system}

\author{Takayuki Myo\footnote{myo@ge.oit.ac.jp,~~takayuki.myo@oit.ac.jp}}
\affiliation{General Education, Faculty of Engineering, Osaka Institute of Technology, Osaka, Osaka 535-8585, Japan}
\affiliation{Research Center for Nuclear Physics (RCNP), Osaka University, Ibaraki, Osaka 567-0047, Japan}

\author{Hiroshi Toki\footnote{toki@rcnp.osaka-u.ac.jp}}
\affiliation{Research Center for Nuclear Physics (RCNP), Osaka University, Ibaraki, Osaka 567-0047, Japan}

\author{Kiyomi Ikeda\footnote{k-ikeda@postman.riken.go.jp}}
\affiliation{RIKEN Nishina Center, Wako, Saitama 351-0198, Japan}

\author{Hisashi Horiuchi\footnote{horiuchi@rcnp.osaka-u.ac.jp}}
\affiliation{Research Center for Nuclear Physics (RCNP), Osaka University, Ibaraki, Osaka 567-0047, Japan}

\author{Tadahiro Suhara\footnote{suhara@matsue-ct.ac.jp}}
\affiliation{Matsue College of Technology, Matsue 690-8518, Japan}

\date{\today}

\begin{abstract}%
We study the tensor-optimized antisymmetrized molecular dynamics (TOAMD) as a successive variational method in many-body systems with strong interaction for nuclei. In TOAMD, the correlation functions for the tensor force and the short-range repulsion and their multiples are operated to the AMD state as the variational wave function. 
The total wave function is expressed as the sum of all the components and the variational space can be increased successively with the multiple correlation functions to achieve convergence.
All the necessary matrix elements of many-body operators, consisting of the multiple correlation functions and the Hamiltonian, are expressed analytically using the Gaussian integral formula.  In this paper we show the results of TOAMD with up to the double products of the correlation functions for the $s$-shell nuclei, $^3$H and $^4$He, using the nucleon-nucleon interaction AV8$^\prime$.  It is found that the energies and Hamiltonian components of two nuclei converge rapidly with respect to the multiple of correlation functions.  This result indicates the efficiency of TOAMD for the power series expansion in terms of the tensor and short-range correlation functions.
\end{abstract}


\maketitle 

\section{Introduction} \label{sec:intro}

One of the central issues in nuclear physics is to understand the nuclear structure from the nucleon-nucleon ($NN$) interaction.
The $NN$ interaction has a strong tensor force at long and intermediate distances and a strong repulsion at short distance \cite{pieper01,pieper01b}.  
It is important to investigate the nuclear structure considering the above characteristics of the $NN$ interaction.

The origin of the tensor force is the one-pion exchange interaction, which brings the high-momentum components of nucleon motion in nuclei. 
It is necessary to treat the high-momentum components induced by the tensor force in the nuclear wave function. 
The tensor force also produces the characteristic $D$-wave state of a nucleon pair in nuclei, which comes from the strong $S$-$D$ coupling
of the tensor force. 
This $D$-wave state is spatially compact as compared with the $S$-wave state due to the high-momentum component of the tensor correlation \cite{ikeda10}. 
The high-momentum component in nuclei coming from the tensor correlation has been investigated experimentally with the ($p$,$d$) reaction \cite{ong13}.

So far, we have described the tensor correlation with high-momentum components on the shell model basis, which we name "tensor-optimized shell model" (TOSM) \cite{myo05,myo09,myo11}.
In TOSM, we fully optimize the two-particle two-hole (2p2h) states in the wave function. There is no truncation for the particle states in TOSM.
In particular, the spatial shrinkage of the particle states is essential to achieve convergence of the contributions of tensor force.  
This property is related to the inclusion of the spatially compact $D$-wave state with high momentum in the wave function.

The clustering of nucleons is one of the important aspects in the nuclear structure, such as the two-$\alpha$ state in $^8$Be and the Hoyle state in $^{12}$C as the triple-$\alpha$ state \cite{ikeda68,horiuchi12}.
Those clustering states can coexist with shell model-like states in a nucleus such as $^{12}$C, the ground state of which is considered to be the shell model-like state.
Theoretically, it is generally difficult to describe the clustering states in the shell model type approach, while the shell model-like states are fairly described \cite{myo11,barrett13,myo14}.
It is also known that the $\alpha$ cluster itself contains the large contribution of the tensor force \cite{myo09,kamada01}.
The relation between the $NN$ interaction and the coexistence of the clustering states and the shell model-like states is unclear. 

It is important to understand the nuclear clustering phenomena from the viewpoint of the $NN$ interaction and the tensor force.
One of the theoretical approaches to describe the nuclear clustering is the antisymmetrized molecular dynamics (AMD) \cite{kanada03,kanada12}.
The AMD wave function consists of the Gaussian wave packet for each nucleon, which is suitable to express the formation of cluster with spatial localization of some of nucleons in a nucleus. 
So far, AMD has shown the successful results in the description of various clustering states in finite nuclei from light mass to medium mass region \cite{horiuchi12}.
However, this model cannot treat the tensor force and/or short-range repulsion, and 
it is necessary for the AMD analysis to rely on the effective interaction of mild central force and $LS$ force without the tensor force.

For the clustering description of nuclei based on the $NN$ interaction, the unitary correlation operator method (UCOM) has been developed to treat the short-range and tensor correlations \cite{feldmeier98,neff04}. 
Using the Fermionic molecular dynamics (FMD) with UCOM, they have discussed the clustering phenomena \cite{neff11}. 
In UCOM, the unitary-transformed Hamiltonian is truncated up to the two-body operator, while the exact transformation produces many-body operators. 
This truncation seems reasonable for short-range repulsion because of the short-range character, but tensor force has a long-range character and many-body operators should be important for the tensor correlation to work correctly. 
The many-body operators are also important for the consistent treatment of the variational principle starting from the $NN$ interaction.

In our study of TOSM, only the short-range part of UCOM is adopted to describe the short-range correlation in the shell-model type basis states,
while the tensor correlation is explicitly treated using the full 2p2h excitation in the wave function.
The method of TOSM+UCOM nicely works to describe the shell model-like states with the correct order of the energy level in the $p$-shell nuclei,
while the $\alpha$ clustering states such as those in $^8$Be and $^{12}$C are difficult to describe quantitatively \cite{myo14,myo12}.

Toward the nuclear clustering description from the $NN$ interaction, we have proposed a new variational theory \cite{myo15}.
We employ the antisymmetrized molecular dynamics (AMD) \cite{kanada03,kanada12} as the basis state.
We introduce two-kinds of correlation functions of the tensor-operator type for the tensor force and the central-operator type for the short-range repulsion. 
This physical concept is similar to UCOM \cite{neff04}.
The correlation functions are multiplied to the AMD wave function as the correlated basis states and superposed with the AMD wave function.
We name this framework ``tensor-optimized antisymmetrized molecular dynamics'' (TOAMD) \cite{myo15}.
In TOAMD, the products of the Hamiltonian and correlation functions become the series of the many-body operators, 
which are exactly treated using the cluster expansion.
We take all the necessary many-body operators without any truncation, which enable us to determine the correlation functions variationally.
The formulation of TOAMD is common for all nuclei with various mass numbers.
The scheme of TOAMD is extendable by taking the series of the multiple product of correlation functions as the power expansion.
This is done systematically and successively in TOAMD and necessary formula are published~\cite{myo15}.

In this paper, we take up to the double products of correlation functions of tensor and short-range types, and investigate the convergence of the solutions with respect to the multiples of correlation functions and discuss the role of each term.
To demonstrate the new successive variational method, we take the $s$-shell nuclei, $^3$H and $^4$He, using the AV8$^\prime$ $NN$ interaction.

\section{Tensor-optimized antisymmetrized molecular dynamics (TOAMD)} \label{sec:TOAMD}

We explain the basic formulation of TOAMD, while all the details are given in Ref. \cite{myo15}.
We start from the AMD wave function, which is expressed by using the Slater determinant of the Gaussian wave packets of nucleons with mass number $A$.
The AMD wave function $\Phi_{\rm AMD}$ is explicitly given as:
\begin{eqnarray}
\Phi_{\rm AMD}
&=& \frac{1}{\sqrt{A!}} {\rm det} \left\{ \prod_{i=1}^A \phi_i \right\}~,
\label{eq:AMD}
\\
\phi(\vec r)&=&\left(\frac{2\nu}{\pi}\right)^{3/4} e^{-\nu(\vec r-\vec D)^2} \chi_{\sigma} \chi_{\tau}~.
\label{eq:Gauss}
\end{eqnarray}
The single-nucleon wave function $\phi(\vec r)$ consists of a Gaussian wave packet with a range parameter $\nu$ and a centroid position $\vec D$, 
the spin part $\chi_{\sigma}$ and isospin part $\chi_{\tau}$.
In this study of $s$-shell nuclei, $\chi_{\sigma}$ is fixed as up or down component and $\chi_{\tau}$ is proton or neutron component.
The range $\nu$ is common for all nucleons and this condition factorizes the center-of-mass wave function from $\Phi_{\rm AMD}$.
The range $\nu$ also contributes to the spatial size of $\Phi_{\rm AMD}$.

In TOAMD we include two-kinds of correlations induced by the tensor force and short-range repulsion, which are difficult to treat in the AMD wave function $\Phi_{\rm AMD}$.
Following the concept given in Ref. \cite{sugie57,nagata59},
we introduce the pair-type correlation functions $F_D$ for tensor force and $F_S$ for short-range repulsion and multiply them to the AMD wave function.  This choice of the TOAMD wave function is motivated by the success of TOSM \cite{myo09,myo14}.
We superpose these components with the original AMD wave function.
Here we define the basic TOAMD wave function as:
\begin{eqnarray}
\Phi_{\rm TOAMD}^{\rm basic}
&=& (1+F_D)(1+F_S) \times\Phi_{\rm AMD}~,
\label{eq:TOAMD}
\\
F_D
&=& \sum_{t=0}^1\sum_{i<j}^A f^{t}_{D}(r_{ij})\,O^t_{ij}\, r_{ij}^2 S_{12}(\hat{r}_{ij})~,
\label{eq:Fd}
\\
F_S
&=& \sum_{t=0}^1\sum_{s=0}^1\sum_{i<j}^A f^{t,s}_{S}(r_{ij})\,O^t_{ij}\,O^s_{ij}~,
\label{eq:Fs}
\end{eqnarray}
with relative coordinate $\vec r_{ij}=\vec r_i - \vec r_j$, $O^t_{ij}=(\vec \tau_i\cdot \vec \tau_j)^t$ and $O^s_{ij}=(\vec\sigma_i\cdot \vec\sigma_j)^s$.
Here $t$ and $s$ represent the isospin and spin channel of a pair, respectively.
The correlation functions $F_D$ and $F_S$ affect only the relative motion of nucleon pairs in $\Phi_{\rm AMD}$, and do not excite the center-of-mass motion.  The center-of-mass motion is completely removed in TOAMD.
The function $F_D$ induces the relative $D$-wave transition via the tensor operator $S_{12}$:
\begin{eqnarray}
S_{12}(\hat r_{ij})=3(\vec \sigma_i\cdot \hat r_{ij})(\vec \sigma_j \cdot \hat r_{ij})-\vec \sigma_i \cdot \vec \sigma_j~.
\end{eqnarray}
The functions $F_D$ and $F_S$ are scalar operator and do not change the total angular-momentum state of $\Phi_{\rm AMD}$.
In general, two functions $F_D$ and $F_S$ are not commutable.
Physically, the functions $F_D$ and $F_S$ can excite two nucleons in the AMD state to the high-momentum region corresponding to the 2p2h excitation in the shell model. 
This formulation of TOAMD is independent of the mass number $A$ and commonly used for all nuclei.  

We state here the essential difference of TOAMD from the Green's function Monte-Carlo (GFMC) method \cite{pieper01}.
In the GFMC method, the standard concept of correlation function is used, where it is expressed by a product: 
\beq
F_S^{\rm GFMC}&=&\prod_{i<j}^A\left(1+\sum_{t=0}^1\sum_{s=0}^1 f^{t,s}_{S}(r_{ij})\,O^t_{ij}\,O^s_{ij}\right)~.
\label{eq:product}
\eeq
for the short-range correlation as an example. 
This makes the calculation of matrix elements complicated, since full $A$-body operators should be calculated with the correlations of every nucleon pair.
On the other hand, the basic TOAMD wave function given in Eq.~(\ref{eq:TOAMD}) uses the lowest order term of the standard correlation function.
This truncation of the correlation operators makes the calculation easier and more systematical.
In TOAMD, we can increase the necessary terms of explicit calculations with the multiple number of correlation functions. 
For the short-range correlation, the TOAMD wave function can be extended as:
\beq
\Phi_{\rm TOAMD}^{\rm short} =  (1+F_S+F_S^2+ \cdots ) \times \Phi_{\rm AMD}~.
\label{eq:expansion}
\eeq
We can examine the convergence of the solutions with the power of the correlation functions step by step.
It is noted that the term $F_S^2$ in Eq.~(\ref{eq:expansion}) has the component of $(f_S^{t,s}(r_{ij}))^2$ which does not appear in Eq.~(\ref{eq:product}).
As for the tensor correlation $F_D$, the same discussion holds.

In the present study of TOAMD, we include up to the double products of the correlation functions consisting of $F_S$ and $F_D$.
The effort of calculating the $F_D F_S$ term in Eq.\,(\ref{eq:TOAMD}) is the same as calculating further the $F_S F_S$, and $F_D F_D$ terms.  This consideration brought us to an idea to use the TOAMD as a successive variational method with respect to the multiple of correlation functions.
Writing all the possible double products of $F_D$ and $F_S$, we come up with the next order of the TOAMD wave function as:
\begin{eqnarray}
\Phi_{\rm TOAMD} 
&=& (1 + F_S + F_D + F_S F_S + F_S F_D + F_D F_S
\nonumber\\
&&  + ~F_D F_D )\times\Phi_{\rm AMD}~.
\label{eq:TOAMD2}
\end{eqnarray}
It is noted that the correlation functions in each term in Eq.\,(\ref{eq:TOAMD2}) are determined independently. 
This means that each correlation function can be different in TOAMD.
All the matrix elements are summed up for the final results.
Hereafter, we call $F_D$ and $F_S$ single correlation functions and their double products are double correlation functions.

The total energy in TOAMD is given as:
\begin{eqnarray}
E_{\rm TOAMD}
&=&\frac{\langle\Phi_{\rm TOAMD} |H|\Phi_{\rm TOAMD}\rangle}{\langle\Phi_{\rm TOAMD} |\Phi_{\rm TOAMD}\rangle}
\nonumber\\
&=&\frac{\langle\Phi_{\rm AMD} | \tilde{H} |\Phi_{\rm AMD}\rangle}{\langle\Phi_{\rm AMD} | \tilde{N} |\Phi_{\rm AMD}\rangle}.
\label{eq:E_TOAMD}
\end{eqnarray}
We calculate the matrix elements of the correlated Hamiltonian $\tilde{H}$ and the correlated norm $\tilde{N}$ with the AMD wave function,
where $\tilde{H}$ and $\tilde{N}$ include the products of correlation functions, such as $F_D^\dagger H F_D$ and $F_D^\dagger F_D$. 
These correlated operators become the series of many-body operators according to the particle index of each operator.
In the case of the two-body interaction $V$, $F_D^\dagger V F_D$ is expanded from two-body to six-body operators with various combinations of particle index.
Similarly, $F_D^\dagger F_D^\dagger V F_D F_D$ gives ten-body operators at maximum.
We classify these many-body operators fully in terms of the cluster expansion method, the detailed procedure of which is given in Ref.~\cite{myo15}.
We take the matrix elements of all the resulting many-body operators with the AMD wave function without any truncation.
This treatment is important to retain the variational principle.
The procedure is performed systematically for all the correlated operators with multiple correlation functions in TOAMD.
In general, many-body operators produce larger number of terms of the cluster expansion for larger mass systems and the calculation of their matrix elements becomes much more demanding.

The TOAMD wave function has three-kinds of variational functions, two-kinds of correlation functions $F_D$, $F_S$ and the AMD wave function $\Phi_{\rm AMD}$.
We determine them using the Ritz variational principle with respect to the TOAMD energy $\delta E_{\rm TOAMD}=0$.

The radial forms of $F_D$ and $F_S$ are optimized in each spin-isospin channel to minimize the total energy $E_{\rm TOAMD}$ in Eq.~(\ref{eq:E_TOAMD}).
We use the Gaussian expansion method to express the relative motion of the pair functions $f^{t}_{D}(r)$ in Eq.~(\ref{eq:Fd}) and $f^{t,s}_{S}(r)$ in Eq.~(\ref{eq:Fs}), respectively, which are given as:
\begin{eqnarray}
   f^t_D(r)
&=& \sum_{n=1}^{N_G} C^t_n \exp(-a^t_n r^2)~,
   \label{eq:cr_D}
\\
   f^{t,s}_S(r)
&=& \sum_{n=1}^{N_G} C^{t,s}_n \exp(-a^{t,s}_n r^2)~.
   \label{eq:cr_S}
\end{eqnarray}
Here, $a^t_n$, $a^{t,s}_n$, $C^t_n$ and $C^{t,s}_n$ are variational parameters for Gaussian basis functions. 
We employ the common value for the Gaussian basis number $N_G$, which is taken seven at most until the solutions are converged.
For the ranges $a^t_n$, $a^{t,s}_n$, we search for the optimized values from short to long ranges to express the spatial correlation adequately, which gain the total energy.
The coefficients $C^t_n$ and $C^{t,s}_n$ are linear parameters in the TOAMD wave function and are determined variationally by diagonalization of the Hamiltonian matrix elements.
In the double correlation functions in Eq.~(\ref{eq:TOAMD2}), the products of two Gaussian functions in Eqs.~(\ref{eq:cr_D}) and (\ref{eq:cr_S}) are treated as the basis functions and the products of $C^t_n$ and $C^{t,s}_n$ are variational parameters for these basis functions.
The centroid positions of the Gaussian wave packets, $\{\vec D_i\}$ ($i=1,\cdots,A$) in Eq.~(\ref{eq:Gauss}) for $\Phi_{\rm AMD}$ are determined variationally
by using the cooling method \cite{kanada03,kanada12}.

In the calculation of matrix elements of many-body operators, we express also the $NN$ interaction as a sum of Gaussian functions.
Technically, we adopt the Fourier transformation of the Gaussian form of the correlation functions and the $NN$ interaction into the momentum space \cite{myo15,goto79}.
This transformation decomposes the many-body operators expressed with the various relative coordinates in the exponent into the separable form with respect to the single particle coordinates.
In the momentum space, the matrix elements of the many-body operators in $\tilde{H}$ and $\tilde{N}$ result in the products of single-particle matrix elements with the AMD basis functions,
which is easily calculated even for various combinations of particle index in the operators.
In this study, we use the realistic $NN$ interaction, AV8$^\prime$ \cite{pieper01b} consisting of central, tensor and $LS$ terms. 

\section{Results}

To demonstrate the TOAMD as a successive variational method in many-body systems with strong interaction, we take the $s$-shell nuclei, $^3$H and $^4$He, where a large number of theoretical results in various few body methods are available~\cite{kamada01}.
We discuss first the TOAMD results for $^3$H and $^4$He with the single correlation function as $\Phi_{\rm TOAMD}= (1 + F_S + F_D)~\Phi_{\rm AMD}$.  
In the variation of the AMD wave function, it is found that the centroid positions of nucleons $\vec D$ are zero for all nucleons in both nuclei.
This result indicates that even with the realistic $NN$ interaction, the $s$-wave configurations are favored in the AMD wave function, 
which is equivalent to the shell-model states of $(0s)^3$ for $^3$H and $(0s)^4$ for $^4$He, respectively.

The correlation functions $F_D$ and $F_S$ are optimized for each nuclei.
The energies are $-5.34$ MeV for $^3$H and $-15.68$ MeV for $^4$He, which are underbound in comparison with the GFMC results; $-7.76$ MeV for $^3$H and $-25.93$ MeV for $^4$He.  
Here, the range parameters in $\Phi_{\rm AMD}$ are $\nu=0.14$ fm$^{-2}$ for $^3$H and $\nu=0.17$ fm$^{-2}$ for $^4$He, which are a little bit different from the values of TOAMD optimized with double correlation functions to be discussed later.
It is noted that both of the tensor and short-range correlations are necessary to make nuclei bound like in the case of the deuteron~\cite{ikeda10}.

\begin{figure}[t]
\centering
\includegraphics[width=8.5cm,clip]{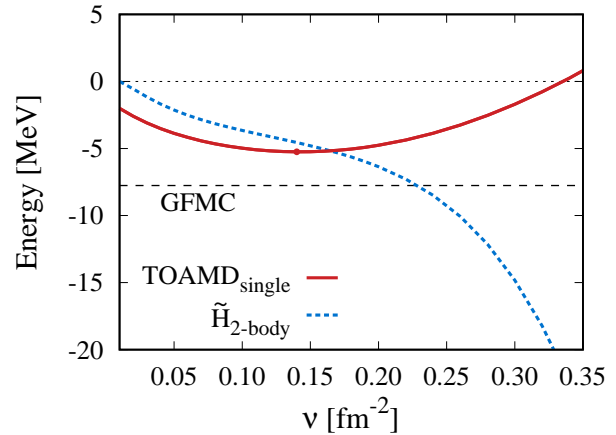}
\caption{Energy surface of $^3$H with AV8$^\prime$ as function of the range parameter $\nu$ in TOAMD with single correlation function (solid line). 
Dotted line represents the results with up to the two-body operators in the correlated Hamiltonian $\tilde{H}$. Dashed line represents the GFMC result.}
\label{fig:ene_3H_nu}
\end{figure}

In the variational calculation, the many-body operators with up to three-body for $^3$H and up to four-body for $^4$He are to be included in the correlated Hamiltonian $\tilde{H}$ and norm $\tilde{N}$. If we perform a limited calculation with up to the two-body operators in the correlated Hamiltonian $\tilde{H}$, which is the similar treatment as UCOM \cite{neff04},
the energies of two nuclei continue to decrease with respect to large $\nu$ and we do not obtain the physical energy minimum.
In this limited calculation in which higher-body operators than two-body are omitted, the variational principle is not satisfied.
In Fig.~\ref{fig:ene_3H_nu}, we confirm this fact by showing the energy surface of $^3$H as function of the range parameter $\nu$ in TOAMD with single correlation function.
The full treatment of many-body operators in $\tilde{H}$ provides the energy minimum properly as shown by solid line. 
On the other hand, when we omit the three-body operators of $\tilde{H}$, there is no energy minimum in $^3$H as shown by dotted line.
This result indicates the inevitable role of the many-body operators in the correlated Hamiltonian to obtain the energy minimum properly.
The similar result is obtained in the Brueckner-Bethe-Goldstone approach for nuclear matter \cite{fukukawa15}, 
in which the three-body correlation terms induced by the $G$-matrix are shown to be necessary to obtain the proper saturation point for density of nuclear matter.
As seen in Fig.~\ref{fig:ene_3H_nu}, our TOAMD approach is very powerful for studying the important roles of many-body operators 
in finite nuclei because it is the variational calculation.
The detailed analysis of many-body operators will be published in the forthcoming paper.

\begin{figure}[t]
\centering
\includegraphics[width=8.5cm,clip]{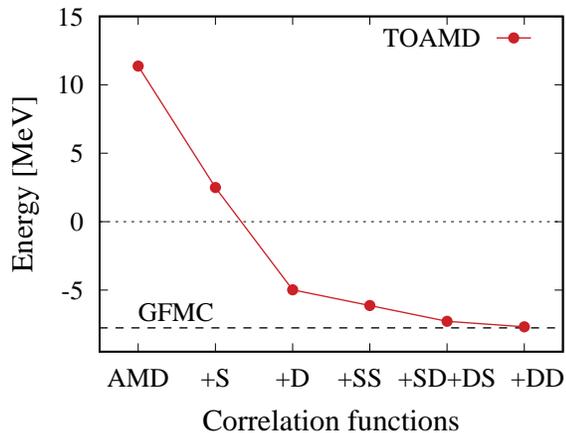}
\caption{Energy convergence of $^3$H with AV8$^\prime$ by adding each term of TOAMD successively. 
Dashed line represents the GFMC result.}
\label{fig:ene_3H_FF}
\end{figure}

Next we include the double correlation functions in TOAMD given in Eq.\,(\ref{eq:TOAMD2}).
We keep the nucleon positions $\vec D=0$ for all nucleons.
In Fig.\,\ref{fig:ene_3H_FF}, we show the energy of $^3$H successively obtained by adding the single and double correlation functions one by one, 
where $\nu=0.095$ fm$^{-2}$ determined variationally in the full calculation.
In each calculation, this value of $\nu$ is common but the correlation functions are optimized independently.
In the figure, the labels D and S indicate $F_D$ and $F_S$, respectively, and ``+S'' is the result obtained with the wave function of $(1+F_S)\times \Phi_{\rm AMD}$. ``+SS'' is the result by adding $F_S F_S$ component as $(1+F_S+F_D+F_SF_S) \times \Phi_{\rm AMD}$ in total.
The components of $F_S F_D$ and $F_D F_S$ are found to give almost identical effect on the solutions, and the combined energy is provided for ``+SD+DS''.
``+DD'' is the full calculation with the double correlation functions in TOAMD.
The final energy up to $F_DF_D$ (+DD) is $-7.68$ MeV, which almost reproduces the GFMC results within 80 keV.

\begin{figure}[t]
\centering
\includegraphics[width=8.5cm,clip]{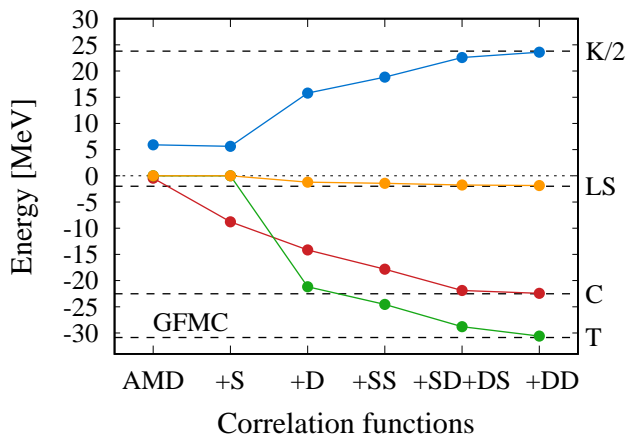}
\caption{Hamiltonian components of $^3$H by adding each term of TOAMD successively. Dashed lines represent the GFMC results for each component.
For the kinetic energy, a half value is shown with the symbol ``K/2''. The symbols C, T and LS indicate the central, tensor and $LS$ forces, respectively.}
\label{fig:ham_3H_FF}
\end{figure}

Figure \ref{fig:ene_3H_FF} shows clearly the effect of each term on the energy.  
At the level of pure AMD the nucleus is not bound, while the addition of $F_S$ decreases the energy and further addition of $F_D$ finally brings the nucleus bound. 
Adding further the double correlation functions step by step, the energy curve shows converging behavior and eventually converges to the GFMC value. 
In this sense we are able to say that the TOAMD is a successive variational method in many-body systems with strong interaction.

\begin{figure}[t]
\centering
\includegraphics[width=8.5cm,clip]{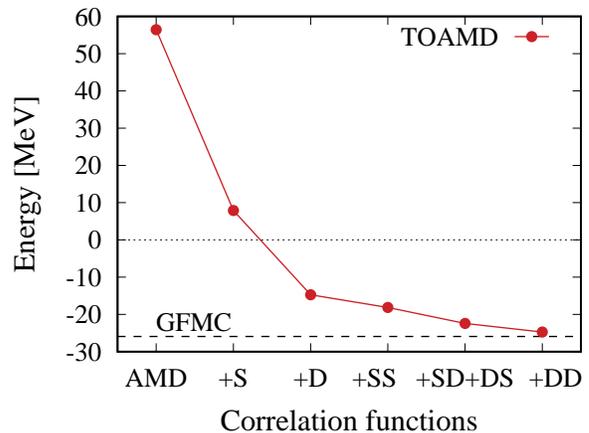}
\caption{Energy convergence of $^4$He with AV8$^\prime$ by adding each term of TOAMD successively.  Dashed line represents the GFMC result.}
\label{fig:ene_4He_FF}
\end{figure}
\begin{figure}[t]
\centering
\includegraphics[width=8.5cm,clip]{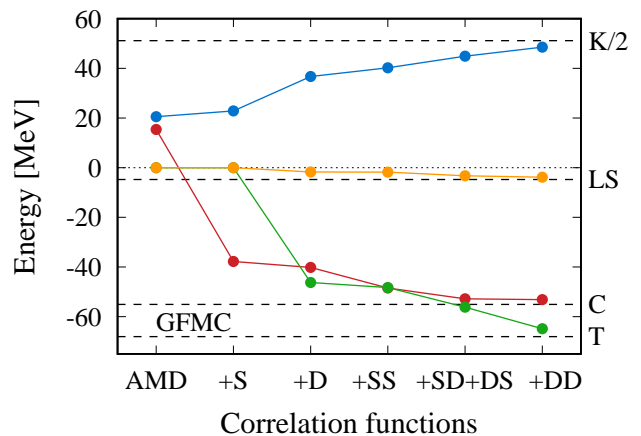}
\caption{Hamiltonian components of $^4$He by adding each term of TOAMD successively. Dashed lines represent the GFMC results for each component.
For the kinetic energy, half value is shown with the symbol ``K/2''. The symbols C, T, LS indicate the central, tensor and $LS$ forces, respectively.}
\label{fig:ham_4He_FF}
\end{figure}

We provide the contributions of each term in the Hamiltonian of $^3$H : the kinetic energy (K), central (C), tensor (T) and $LS$ (LS) forces in Fig.\,\ref{fig:ham_3H_FF}. 
We see how each term in the Hamiltonian changes as the correlation functions are added successively. 
The contributions of the tensor and $LS$ forces are found to have the values after adding the tensor correlation (+D).
Again each term converges to the corresponding results of the GFMC calculation.
The matter radius is obtained as 1.746 fm. 
From these results, we can learn the accuracy and power of TOAMD and the efficiency of correlation functions $F_D$ and $F_S$ to treat the $NN$ interaction.
Good reproduction of Hamiltonian components of $^3$H including the $LS$ force indicates that the tensor and short-range correlations are essential and sufficient in the description of the $^3$H wave function.  This is similar to the case of the deuteron, 
where the $D$-wave component brought by the tensor force provides the $LS$ energy.
In Fig. \ref{fig:ham_3H_FF}, it is found that the ''AMD+S'' calculation does not provide the enhancement of the kinetic energy from the AMD result in spite of the inclusion of the short-range correlation.
This is because $F_S$ can represent not only the short-range character but also the intermediate- and long-range ones.

We discuss now the case of $^4$He. Figure \ref{fig:ene_4He_FF} shows the total energy obtained by adding successively the correlation functions in TOAMD wave function in Eq.~(\ref{eq:TOAMD2}), where the range parameter of the AMD wave function is $\nu=0.22$ fm$^{-2}$. We see quite a similar behavior of the total energy as that of $^3$H. 
We see a good convergence pattern with the correlation functions. With full components up to $F_D F_D$, the energy is obtained as  $-24.74$ MeV in TOAMD. 
The energy difference between TOAMD and GFMC is about 1.2 MeV.

Figure \ref{fig:ham_4He_FF} shows the Hamiltonian components.  Each component as kinetic energy (K), central (C), tensor (T) and $LS$ (LS) forces show gradual convergence with successive addition of the multiple of the correlation functions.  These energies deviate slightly from the GFMC results. 
The matter radius is obtained as 1.497 fm, which is very close to the GFMC value of 1.490 fm. 

We compare the present results of $^4$He with TOSM using short-range UCOM \cite{myo09}, which gives the energy of $-22.30$ MeV.
Regarding the short-range UCOM as $1+F_S$ and the 2p2h excitations in TOSM as the role of $F_S$ or $F_D$ in TOAMD, TOSM corresponds to TOAMD without the $F_D F_D$ term, 
which gives the energy of $^4$He as $-22.44$ MeV.
This value is close to the TOSM result. In this sense, TOAMD includes the tensor correlation more than that of TOSM, owing to the $F_D F_D$ term. 

In TOAMD, the correlation functions $F_S$ and $F_D$ are optimized independently in each term of Eq.~(\ref{eq:TOAMD2}).
It is interesting to see this effect on the energy, and we perform the following calculation.
First, $F_S$ and $F_D$  are determined in the single correlation function of TOAMD as $(1+F_S+F_D) \times \Phi_{\rm AMD}$.
Second, keeping the functional form of $F_S$ and $F_D$ with Gaussian expansion, we perform the calculation including double correlation functions,
where the weights of the double correlation functions are variational parameters.
This calculation provides the energies of $^3$H and $^4$He as $-6.26$ MeV and $-22.40$ MeV, respectively.
The energy loss from the full calculation is 1.44 MeV for $^3$H and 2.34 MeV for $^4$He.
These amounts indicate the importance of the independent optimization of the correlation functions in each term of TOAMD, contributing to the rapid energy convergence.

From the numerical results of $^3$H and $^4$He shown in Figs.\,\ref{fig:ene_3H_FF} and \ref{fig:ene_4He_FF}, it is found that the convergence of energies with respect to the correlation functions is rapid in TOAMD. 
This fact indicates the validity of the present expansion of the wave function in power of the tensor and short-range correlations.
Considering the convergence of the solutions in the present analysis,
we can expect that we reach the precise energy further by increasing the multiples of correlation functions to the next order.
This extension of more correlation functions is handled systematically in TOAMD by taking all kinds of many-body operators emerging from the correlated Hamiltonian and norm.
The next order is the triple products of the correlation functions consisting of $F_D$ and $F_S$.
We can put a priority on the basis functions involving the tensor and short-correlations at the same time, 
which leads to the $F_D F_D F_S$ and $F_D F_S F_S$ products in TOAMD.
Increasing the power of correlation functions, 
the analysis of the results at each power is physically meaningful to get a knowledge of the role of correlation functions.  
This successive variational method in many-body systems with strong interaction is important when we calculate heavier nuclei, since we are able to see how we approach the convergence.

One of the advantages of TOAMD is the clustering description based on the AMD basis states.
It is interesting to consider the system consisting of several clusters such as $^8$Be with two $^4$He.
It is found that each $^4$He nucleus needs the double products of the correlation functions to get the sufficient binding energy.
This fact naively suggests that the spatially separated two-$^4$He state will need fourth power of the correlation functions totally in TOAMD.
It is an interesting problem how the cluster states are described in TOAMD with the increase of the power of correlation functions.

\section{Summary}
We have developed a new variational theory ``tensor-optimized antisymmetrized molecular dynamics'' (TOAMD) 
to describe the nuclear structure using the nucleon-nucleon ($NN$) interaction, in particular, toward the nuclear clustering description.
In TOAMD, the tensor- and central-type correlation functions are introduced considering the characteristics of the $NN$ interaction.
These correlation functions are multiplied to the AMD wave function to express the effects of tensor force and short-range repulsion explicitly and the correlated basis states are superposed with the AMD wave function.
This scheme of TOAMD is independent of the mass number and extendable by increasing the power of the multiple products of the correlation functions successively.
In the calculation of matrix elements, the products of the Hamiltonian and the correlation functions produce the many-body operators in principle,
which are exactly treated without any truncation using the cluster expansion. This is important to keep the variational principle starting from the $NN$ interaction.
The genuine three-nucleon interaction such as Fujita-Miyazawa type can be tractable in TOAMD in the same manner as treating many-body operators. 

In this study, we take up to the double products of correlation functions in TOAMD and show the effect of each term step by step successively. 
Using the AV8$^\prime$ $NN$ interaction, we show the efficiency of TOAMD in the description of $s$-shell nuclei.
The TOAMD results reproduce the $^3$H energy and provide a good binding energy for $^4$He in the scheme of the double correlation functions.
The radius is also reproduced for two nuclei.
These results indicate that the essential correlations induced by the $NN$ interaction is sufficiently included in the present order of TOAMD for $s$-shell nuclei, which is physically meaningful and also useful when we treat the system of heavier mass nuclei.  In addition, the convergence of the solutions with respect to the multiples of correlation functions is rapid.
This indicates the validity of the present expansion approach of TOAMD.
We believe that it is important to obtain the precise energy for $^4$He, since it is a building block of light nuclei.
In order to reduce the energy difference from the GFMC value, we shall increase the multiples of correlation functions with the triple products such as $F_D F_D F_S$ as the next order.
Based on the success for the $s$-shell nuclei, we shall apply TOAMD to the $p$-shell nuclei with the three-nucleon interaction.

\section*{Acknowledgments}
This work was supported by JSPS KAKENHI Grant Numbers JP15K05091, JP15K17662, JP16K05351.

\section*{References}
\def\JL#1#2#3#4{ {{\rm #1}} \textbf{#2}, #4 (#3)}  
\nc{\PR}[3]     {\JL{Phys. Rev.}{#1}{#2}{#3}}
\nc{\PRC}[3]    {\JL{Phys. Rev.~C}{#1}{#2}{#3}}
\nc{\PRA}[3]    {\JL{Phys. Rev.~A}{#1}{#2}{#3}}
\nc{\PRL}[3]    {\JL{Phys. Rev. Lett.}{#1}{#2}{#3}}
\nc{\NP}[3]     {\JL{Nucl. Phys.}{#1}{#2}{#3}}
\nc{\NPA}[3]    {\JL{Nucl. Phys.}{A#1}{#2}{#3}}
\nc{\PL}[3]     {\JL{Phys. Lett.}{#1}{#2}{#3}}
\nc{\PLB}[3]    {\JL{Phys. Lett.~B}{#1}{#2}{#3}}
\nc{\PTP}[3]    {\JL{Prog. Theor. Phys.}{#1}{#2}{#3}}
\nc{\PTPS}[3]   {\JL{Prog. Theor. Phys. Suppl.}{#1}{#2}{#3}}
\nc{\PTEP}[3]   {\JL{Prog. Theor. Exp. Phys.}{#1}{#2}{#3}}
\nc{\PRep}[3]   {\JL{Phys. Rep.}{#1}{#2}{#3}}
\nc{\PPNP}[3]   {\JL{Prog.\ Part.\ Nucl.\ Phys.}{#1}{#2}{#3}}
\nc{\JP}[3]     {\JL{J. of Phys.}{#1}{#2}{#3}}
\nc{\andvol}[3] {{\it ibid.}\JL{}{#1}{#2}{#3}}

\end{document}